\pdfoutput=1

\documentclass[11pt]{article}

\usepackage{EMNLP2023}

\usepackage{times}
\usepackage{latexsym}
\usepackage{xspace, subfigure}
\usepackage{amsmath}
\usepackage{amssymb}

\usepackage{hyperref}
\usepackage{adjustbox}
\usepackage{array}
\usepackage{balance}
\usepackage{multirow}
\usepackage{babel, caption}
\usepackage[T1]{fontenc}
\usepackage[utf8]{inputenc}
\usepackage{microtype}
\usepackage{inconsolata}
\usepackage{titlesec}

\titlespacing\section{0pt}{6pt plus 8pt minus 4pt}{0pt plus 2pt minus 3pt}
\titlespacing\subsection{0pt}{4pt plus 6pt minus 2pt}{0pt plus 2pt minus 3pt}
\titlespacing\subsubsection{0pt}{12pt plus 4pt minus 2pt}{0pt plus 2pt minus 3pt}

\title{An Empirical Study of Using ChatGPT for Fact Verification Task}

\begin{document}

\author{Mohna Chakraborty, Adithya Kulkarni, \and Qi Li \\ Department of Computer Science, Iowa State University \\ \{mohnac, aditkulk, qli\}@iastate.edu}
\maketitle

\begin{abstract}

ChatGPT has recently emerged as a powerful tool for performing diverse NLP tasks. However, ChatGPT has been criticized for generating nonfactual responses, raising concerns about its usability for sensitive tasks like fact verification. This study investigates three key research questions: (1) Can ChatGPT be used for fact verification tasks? (2) What are different prompts performance using ChatGPT for fact verification tasks? (3) For the best-performing prompt, what common mistakes does ChatGPT make? Specifically, this study focuses on conducting a comprehensive and systematic analysis by designing and comparing the performance of three different prompts for fact verification tasks on the benchmark FEVER dataset using ChatGPT.

\end{abstract}
\section{Introduction}

With recent advancements in scalable attention-based architectures \cite{vaswani2017attention}, language models encompass phenomenal generalization and exhibit unprecedented performance at generating novel human-like content. One such large language model based on GPT-3.5-TURBO or GPT-4 architecture is ChatGPT \cite{openai2023gpt4}, capable of generating text responses nearly indistinguishable from those written by humans. The generated text reads remarkably convincing, but several factors should be considered while evaluating the model's response. Specifically, we need to check whether the response is opinionated, contains toxic aspects, or is accurate. Recent studies \cite{ray2023chatgpt, zhuo2023exploring} have evaluated and analyzed the ethical aspects of ChatGPT; however, there has yet to be a comprehensive study that investigates the accuracy of ChatGPT for the fact verification task. 

Fact verification requires the model to verify a given textual claim based on data from reliable sources. ChatGPT is known to generate convincing responses; however, the response may not be accurate. Due to the sensitive nature of the fact verification task, it is quintessential to empirically validate the accuracy of the response, including the common mistakes made by ChatGPT. In this study, we conduct a systematic empirical analysis to validate the usability of ChatGPT for the fact verification task. 

Another observation made by many users is that the performance of ChatGPT is sensitive to the choice of prompts~\cite{li2023preliminary}. A well-designed prompt can significantly improve the quality and relevance of ChatGPT's output, while a poorly crafted one may lead to incorrect responses. Prompt designing of ChatGPT for the fact verification task requires understanding the task objectives, the output format, and the strengths and limitations of ChatGPT. To design the prompt, we can directly query ChatGPT for the prompt or design it manually. In this study, we design and compare the performance of ChatGPT on the fact verification task with three different prompts. Specifically, we aim to understand how well the prompt suggested by ChatGPT works in practice and if there is a better prompt that can achieve better performance. 

In summary, this study investigates three key research questions: \textbf{RQ1:} Can ChatGPT be used for fact verification tasks? \textbf{RQ2:} What are the different prompts performance using ChatGPT for fact verification tasks? \textbf{RQ3:} For the best-performing prompt, what common mistakes does ChatGPT make?

We conduct experiments on the development and test splits of the non-binary FEVER 1.0 benchmark dataset to answer the research questions. For the experiments, we design three non-binary prompts\footnote{prompts that require ChatGPT to provide non-binary response.}, one suggested by ChatGPT, and two handcrafted prompts. We comprehensively analyze the performance of ChatGPT on the three designed prompts and systematically verify the common errors made by ChatGPT on the best-performing prompt. Furthermore, we also compare the designed prompts with a binary prompt designed by \citet{hoes2023using}.

\section{Related Works}

Prompts are widely used to employ LLMs for various NLP tasks. In this section, we discuss the related studies that explored prompt design for LLMs like ChatGPT.

Prompts are designed either manually \cite{schick2021exploiting} or automatically \cite{shin2020autoprompt, gao2021making}. Several prompt engineering methods are proposed to guide the model to generate better intermediate reasoning explanations to solve a complex task. For instance, \citet{cai2023large} proposed the Chain-of-Thought (CoT) method to improve the reasoning abilities of LLMs by decomposing complicated problems into multiple intermediate steps and solving them individually, and using their results to solve the final problem.

Specifically for ChatGPT, \citet{han2021ptr} proposed strategies for designing prompts for classification tasks. \citet{sobania2023analysis, xia2023conversational} examined prompts for software bug fixing. \citet{wang2023chatgpt} utilized ChatGPT to generate effective boolean prompts for systematic review literature search. In several domains, various studies examined the positive usage of ChatGPT results. Specifically, \citet{sallam2023chatgpt, gilson2023does} observed surprisingly good performance of ChatGPT in medical licensing exams. \citet{choi2023chatgpt} and \citet{frieder2023mathematical} explored using ChatGPT in Law School and for mathematical reasoning abilities, respectively. 

More related to our work, several studies also consider the security risks and ethics of ChatGPT. To test for severe privacy concerns, \citet{li2023multi} developed a multi-step jailbreaking attack against ChatGPT to steal confidential information. Jailbreak prompts are used to remove the restrictions on ChatGPT and facilitate its usage beyond its intended capacity. \citet{liu2023jailbreaking} reviewed the existing jailbreak prompts and their ability to circumvent the limitations imposed on ChatGPT. \citet{huang2023chatgpt} did an empirical evaluation to study the performance of ChatGPT in classifying and generating explanations for implicit hate speech. \citet{khalil2023will} studied ChatGPT in academic settings from the perspective of academic honesty and plagiarism. \citet{hoes2023using} evaluated the performance of ChatGPT at classifying verified statements by designing a binary prompt for PolitiFact dataset \cite{misra2022politifact}.

\section{Prompts for Fact Verification Task}
\label{prompt_design}

\subsection{FEVER Challenge}

The Fact Extraction and VERification (FEVER) Shared Task~\cite{thorne2018fever} conducted in 2018 challenged participants to classify if the human-written factoid claims can be SUPPORTED or REFUTED using the evidence retrieved from Wikipedia June 2017 dump\footnote{https://fever.ai/dataset/fever.html}. If sufficient evidence is absent in Wikipedia, such claims are labeled Not Enough Information (NEI). 

\subsection{Prompt Designing}

We prompted ChatGPT with the following three prompts: 

\textbf{Prompt-1:} The claim is "<claim>". Select whether the claim is (a.) – Support, (b.) – Refute, (c.) – NEI?

\textbf{Prompt-2:} The claim is "<claim>". Provide evidence for the claim, and based on the evidence, select whether the claim is (a.) - Support, (b.) - Refute, (c.) - NEI?

\textbf{Prompt-3:} The claim is "<claim>". Can you generate 5 prompts for this claim and the answers for each of the prompts? For each of the answers, select whether that prompt is (a.) - Support, (b.) – Refute, (c.) – NEI? Based on the prompts and the answers provided, select finally whether the claim is (a.) - Support, (b.) – Refute, (c.) – NEI?

\textit{Prompt-1} asks ChatGPT to directly provide the label for the claim. \textit{Prompt-2} asks ChatGPT to provide evidence for the claim along with the label. \textit{Prompt-2} is the suggested prompt by ChatGPT for the fact verification task. Comparing performances of ChatGPT on \textit{Prompt-2} and \textit{Prompt-1} provides insights into the importance of evidence for the claim in ChatGPT's verdict for the claim. 
\textit{Prompt-3} asks ChatGPT to generate more prompts related to the claim and answers them. Based on the answers to the generated prompts, ChatGPT is asked to provide the label for the claim. ChatGPT is known to change its response based on the conversation, and in most scenarios, the change is positive. We utilize this observation to design \textit{Prompt-3}. Empirically, we test if asking ChatGPT to generate multiple prompts improves its performance for the fact verification tasks. For each prompt, we constrain the output to be the labels of the task.

\section{Experiments}

We empirically investigate three key research questions on the benchmark dataset from FEVER 1.0 shared task.

\subsection{Dataset} \label{Dataset}
The experiments\footnote{We will publish ChatGPT response for all the designed prompts along with the claims used for error analysis upon acceptance.} are conducted on the development and test splits\footnote{The experiments are conducted in zero-shot setting; therefore, we do not need train split.} of the benchmark dataset from FEVER 1.0 shared task\footnote{\url{https://fever.ai/dataset/fever.html}} \cite{thorne2018fever}. Table \ref{tab:stat} shows the statistics of the dataset. 

\begin{table}[t]
  \caption{The statistics of the FEVER 1.0 dataset as presented in  \citet{thorne2018fever}}
  \small
  \centering
  \begin{tabular}{c|c|c|l}
  \hline
    \textbf{Split}& \textbf{Support} & \textbf{Refute} & \textbf{NEI}\\
    \hline
    Development & 6,666 & 6,666 & 6,666\\
    \hline
    Test & 6,666 & 6,666 & 6,666\\
  \hline
\end{tabular} 
\label{tab:stat}
\end{table}

\subsection{Experimental Setup and Evaluation}

For ChatGPT, we use APIs to access \textit{gpt-3.5-turbo (version March 23, 2023)} model provided by OpenAI\footnote{https://chat.openai.com/}. For each of the prompts discussed in Section \ref{prompt_design}, we substitute the variable "<claim>" with the claims from the FEVER 1.0 dataset and pass it as input to ChatGPT API to get the predictions. 

We report \textbf{Label Accuracy (LA)}, the official metric for the FEVER 1.0 shared task. For evaluating the development set, we use the official scorer provided by the workshop\footnote{https://github.com/sheffieldnlp/fever-scorer}, and for evaluating the test set, we upload the predictions to Codalab\footnote{https://competitions.codalab.org/competitions/18814}. 
We evaluate the predictions under two settings, \textit{setting-1}, where ChatGPT predictions are evaluated for all three labels of Support, Refute, and NEI, and \textit{setting-2}, where the predictions are evaluated only for two labels of Support, and Refute. NEI label depends on the amount of data accessible to the model. Since we do not know and cannot restrict the data available to ChatGPT, \textit{setting-2} gives us a better understanding of the performance of ChatGPT. The evaluation on Codalab is done for all \textit{3 labels} of Support, Refute, and NEI; therefore, we do not report the test set results for \textit{setting-2}. Note that the prompts discussed in Section \ref{prompt_design} are non-binary, implying that the prompts require ChatGPT to choose from one of the three labels. We use the entire development and test splits of the FEVER 1.0 shared task for \textit{setting-1} evaluation, and we remove the claims with ground truth label as NEI for \textit{setting-2} evaluation.

\subsection{Baseline Methods}
For the evaluation of \textit{setting-2}, we compare the performance of ChatGPT on the three non-binary prompts designed in Section \ref{prompt_design} with the binary prompt designed by HOES~\cite{hoes2023using}. The binary prompt designed by HOES is as follows:

\textbf{HOES-prompt:} Can you fact-check a claim "<claim>" for me? When fact checking use clear language such as `true' and `false', and avoid negations.

The binary prompt has an advantage over the non-binary prompt since ChatGPT should only choose between two labels of `true' (Support) and `false' (Refute), whereas the non-binary prompt requires ChatGPT to choose among the three labels of Support, Refute, and NEI.

\subsection{Results: RQ1}

Table \ref{table: main_table} depicts the performance of ChatGPT for the fact verification task using three different prompts. From the results, we can observe that ChatGPT achieved good performance using \textit{Prompt-3}. Specifically, we can observe that ChatGPT achieved $>72\%$ for both the development and test splits for evaluation \textit{setting-1} and achieved $>85\%$ for the development set for evaluation \textit{setting-2}. The results indicate that ChatGPT can achieve good performance when verifying claims. Therefore, ChatGPT can be a supportive tool for the fact verification task on datasets such as FEVER.

\subsection{Results: RQ2}

\begin{table}[t]
\caption{Results for different prompts.}
\small
\centering
\begin{tabular}{c|c|c|c}
\hline
\textbf{Prompt} & \textbf{Setting} & \textbf{Development} & \textbf{Test} \\
\hline
Prompt-1 & Support/ & 52.04 & 51.09 \\ \cline{1-1}\cline{3-4}
Prompt-2 & Refute/ & 53.00 & 51.53 \\ \cline{1-1}\cline{3-4}
Prompt-3 & NEI & 72.12 & 74.36 \\ \hline \hline
HOES-prompt &  & 80.32 & - \\ \cline{1-1}\cline{3-4}
Prompt-1 & Support/& 55.12 & - \\ \cline{1-1}\cline{3-4}
Prompt-2 &  Refute& 68.80 & - \\ \cline{1-1}\cline{3-4}
Prompt-3 &  & \textbf{85.64} & - \\
\hline
\end{tabular}
\label{table: main_table}
\end{table}

Comparing the performance of ChatGPT on \textit{Prompt-1} and \textit{Prompt-2} in Table \ref{table: main_table}, we observe that both prompts achieve similar performance on development and test splits on evaluation \textit{setting-1}. The results suggest that asking for additional evidence for the claim does not improve the performance of ChatGPT. From the results on evaluation \textit{setting-2}, we can observe that \textit{Prompt-2} outperforms \textit{Prompt-1} for the development set, but the performance is still unsatisfactory. 

From the results of ChatGPT in Table \ref{table: main_table}, we can observe that \textit{Prompt-3} significantly outperforms both \textit{Prompt-1} and \textit{Prompt-2} for both evaluation settings. The results suggest that asking ChatGPT to generate multiple prompts and answer them helps it to analyze the claim from different aspects, thus enabling it to make more guided decisions. Specifically, asking ChatGPT to generate multiple prompts creates a flow similar to a conversation resulting in additional information that ChatGPT utilizes to properly reason the response for a claim. We provide case study examples in Appendix \ref{sec:case_study}.

We can also observe that ChatGPT achieves significantly better performance on non-binary \textit{Prompt-3} compared to binary \textit{HOES-prompt} even though binary prompt has an advantage over non-binary prompt. The results further validate that the designed \textit{Prompt-3} can significantly improve the quality and relevance of ChatGPT's output for fact verification tasks on datasets such as FEVER.

\subsection{Results: RQ3}

To investigate the common mistakes made by ChatGPT for fact verification tasks, we examined the prediction of ChatGPT on \textit{prompt-3}. For the experiment, we randomly sampled $200$ claims for which ChatGPT made mistakes. We categorized these claims into eight categories depending on the type of mistake. Each claim is placed into one of the eight categories; if the claim belongs to more than one category, it is placed into the most prominent category. Among the error categories, the most prominent is \textit{Miscellaneous}, where ChatGPT correctly answers and predicts the labels for each of the five prompts it generates but makes a mistake in the overall prediction ($48\%$). We suspect the primary reason for the error is that ChatGPT could not correctly reason the importance of prompt outputs to the claim. The second most prominent error category is \textit{Misunderstanding the claim}, where ChatGPT failed to correctly understand the context of the claim resulting in an incorrect response ($21\%$). The third most prominent error category is \textit{Factual errors}, where ChatGPT made factual errors in the answers provided for the generated prompts resulting in incorrect judgment ($16\%$). The fourth most prominent error category is \textit{Irrelevant prompts}, where some prompts generated by ChatGPT are irrelevant to the claim, leading to a mistake in the output label ($9\%$).   The remaining four categories that account for less than $8\%$ of the errors are \textit{Additional information}, \textit{Missing knowledge}, \textit{Incorrect answers}, and \textit{Time sensitive claim}. \textit{Incorrect answers} error category contains claims where ChatGPT made non-factual mistakes in the answers for the generated prompts. The errors in the categories of \textit{Additional information}, \textit{Missing knowledge}, and \textit{Time sensitive claim} are not true errors and are caused due to the data available to ChatGPT. The evidence for the claims in the FEVER 1.0 dataset are obtained from Wikipedia's June 2017 dump, whereas ChatGPT may or may not have access to Wikipedia or may have access to additional websites resulting in incorrect responses for these categories. Addressing the evaluation to remove these false errors will not increase ChatGPT performance much. We provide examples for each error category in Appendix \ref{sec:error_examples}.

\begin{figure}
    \centering
    \includegraphics[width=0.49\textwidth]{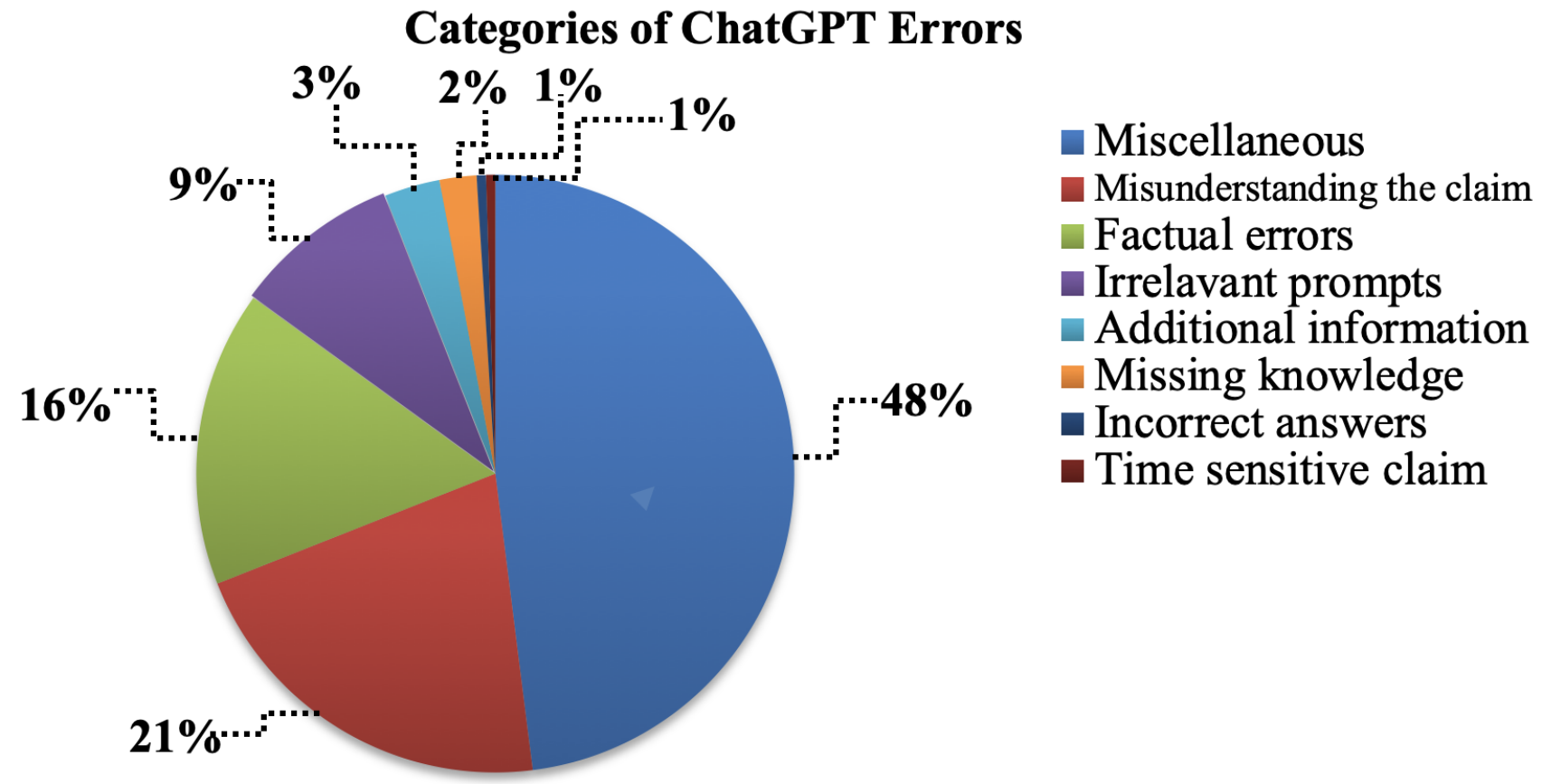}
    \caption{Categories of errors by ChatGPT. The error percentage is rounded off to the nearest integer.}
    \label{fig:fig1}
    \vspace{-0.7cm}
\end{figure}

From the comprehensive error analysis, we find that the primary cause for erroneous responses by ChatGPT lies in incorrect logical reasoning and contextual misunderstanding of the claim. ChatGPT failed to properly reason the generated prompt answers to the claim and failed to better understand the context of the claim and instead focused on the tokens used in the claim.
\section{Conclusion}

In this work, we empirically investigate three key research questions related to the usability of ChatGPT for the fact verification task. We find that ChatGPT can perform well for fact verification tasks on datasets such as FEVER with a correct prompt. From the comprehensive error analysis, we find that near to $70\%$ of the mistakes by ChatGPT are due to incorrect logical reasoning and contextual misunderstanding of the claim. 
\section{Limitations}
In this work, we investigate the usability of ChatGPT for fact-verification tasks on FEVER 1.0 shared task on a specific version \textit{gpt-3.5-turbo (version March 23, 2023)} of the ChatGPT model, and our conclusions cannot be extended to other versions of the model. Furthermore, our conclusions are based on empirical results on FEVER 1.0 shared task and cannot be extended to other fact verification datasets. The ground truth labels of FEVER 1.0 shared task datasets are based on Wikipedia's June 2017 dump, and in our evaluation, we observed that this could lead to some errors. Therefore, our evaluation metric has some limitations.
\section*{Ethics Statement}

We comply with the ACL Code of Ethics. 

\bibliography{anthology,custom}
\bibliographystyle{acl_natbib}
\clearpage
\appendix

\section{Appendix}
\label{sec:appendix}

\subsection{ChatGPT Error Category Examples}
\label{sec:error_examples}
\begin{figure}
    \centering
    \includegraphics[width=0.49\textwidth]{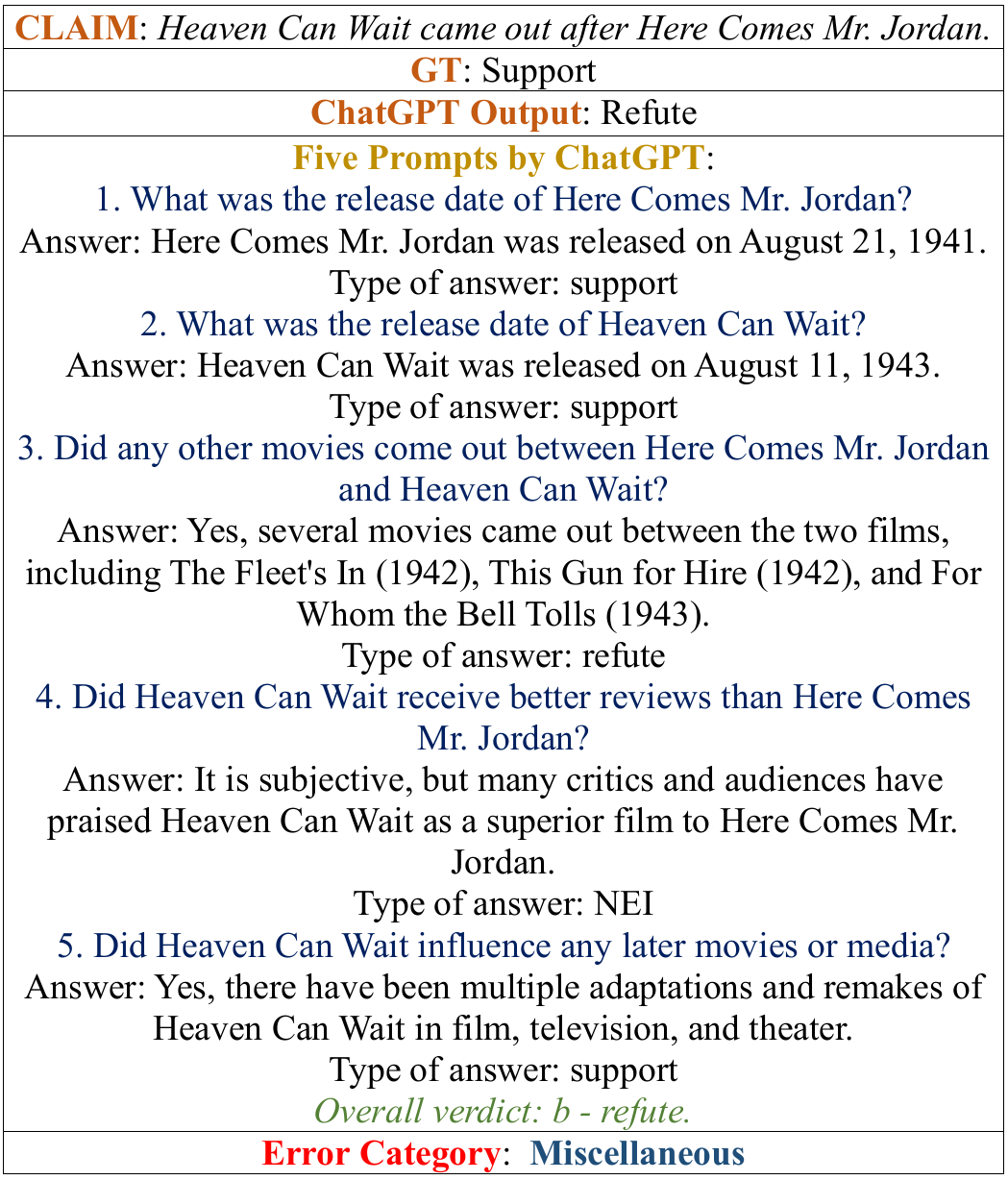}
    \caption{Miscellaneous}
    \label{fig:fig2}
\end{figure}

Figure \ref{fig:fig2} examines the error category \textit{Miscellaneous}, where ChatGPT correctly predicted the label for the claim in the generated five prompts but made a mistake in the overall prediction. We can observe that all the generated prompts and their corresponding answers are correct; however, the error primarily occurred because ChatGPT could not correctly reason the prompt outputs to the claim. 

\begin{figure}
    \centering
    \includegraphics[width=0.49\textwidth]{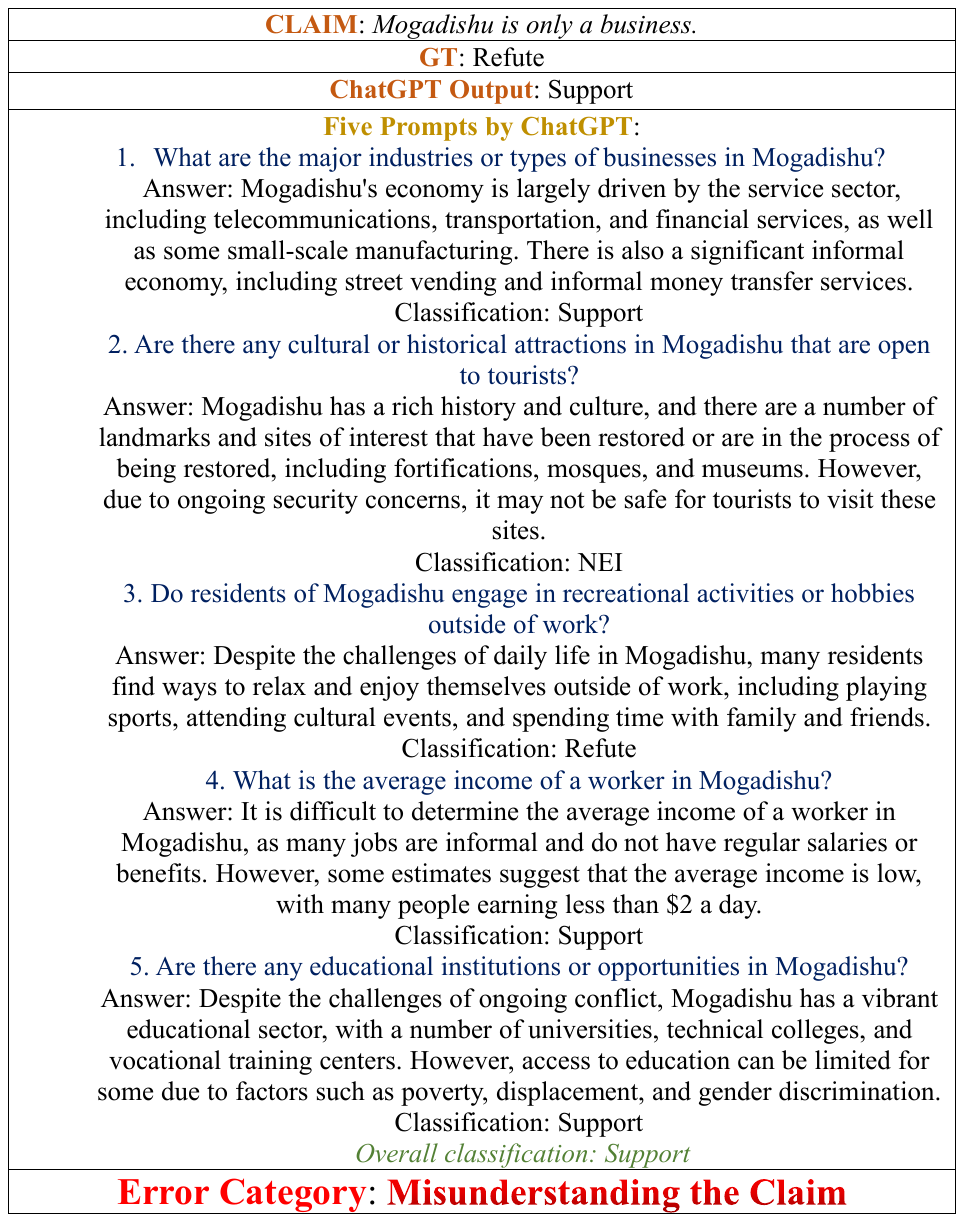}
    \caption{Misunderstanding the claim}
    \label{fig:fig3}
\end{figure}

Figure \ref{fig:fig3} examines the error category \textit{Misunderstanding the claim}, where ChatGPT failed to correctly understand the claim. Specifically, we can observe that the prompts generated by ChatGPT did not understand the context of the claim and focussed only on the words used in the claim. ChatGPT needs to understand the claim's intent about Mogadishu being a city, not just a business, and accordingly generate prompts that align with the claim's intent. ChatGPT failed to understand the context of the claim, thereby misunderstanding the claim.

\begin{figure}
    \centering
    \includegraphics[width=0.49\textwidth]{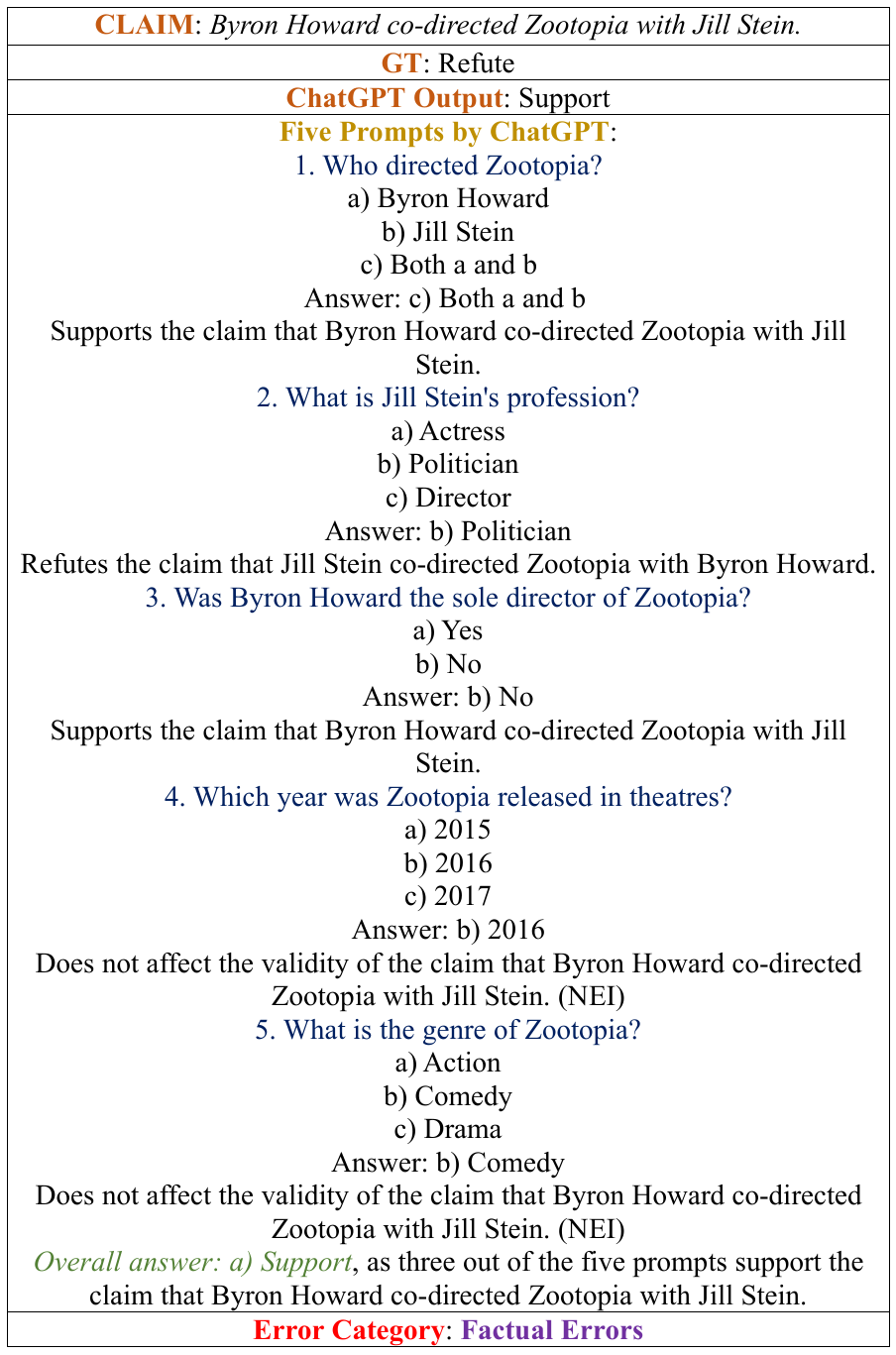}
    \caption{Factual Errors}
    \label{fig:fig4}
\end{figure}

Figure \ref{fig:fig4} examines the error category \textit{Factual errors}, where ChatGPT made factual errors for the answers provided for the generated prompts. Though the movie Zootopia is indeed co-directed; however, Jill Stein is not the co-director. Three of the five prompts supported the claim that Byron Howard co-directed Zootopia with Jill Stein, causing the ChatGPT response to be factually incorrect.

\begin{figure}
    \centering
    \includegraphics[width=0.49\textwidth]{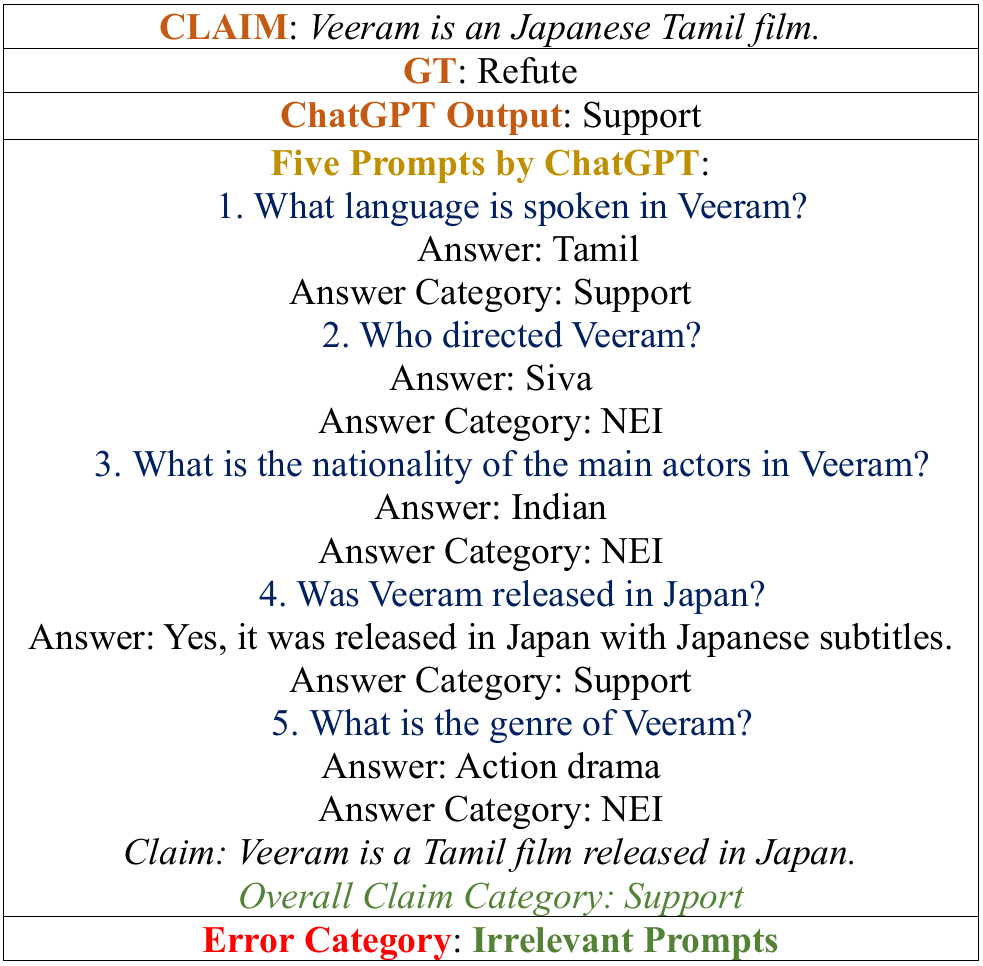}
    \caption{Irrelevant Prompts}
    \label{fig:fig5}
\end{figure}

Figure \ref{fig:fig5} examines the error category \textit{Irrelevant prompts}, where some prompts generated by ChatGPT are irrelevant to the claim. Specifically, we can observe that the prompts generated by ChatGPT do not understand the context in which the word "Japanese" has been used in the claim and generate some irrelevant prompts. The only prompt generated with the word Japan was "Was Veeram released in Japan," whose objective differs from the claim's purpose. A correct understanding of the context will guide ChatGPT to generate prompts such as "Is Veeram a Japanese film?". 

\begin{figure}
    \centering
    \includegraphics[width=0.49\textwidth]{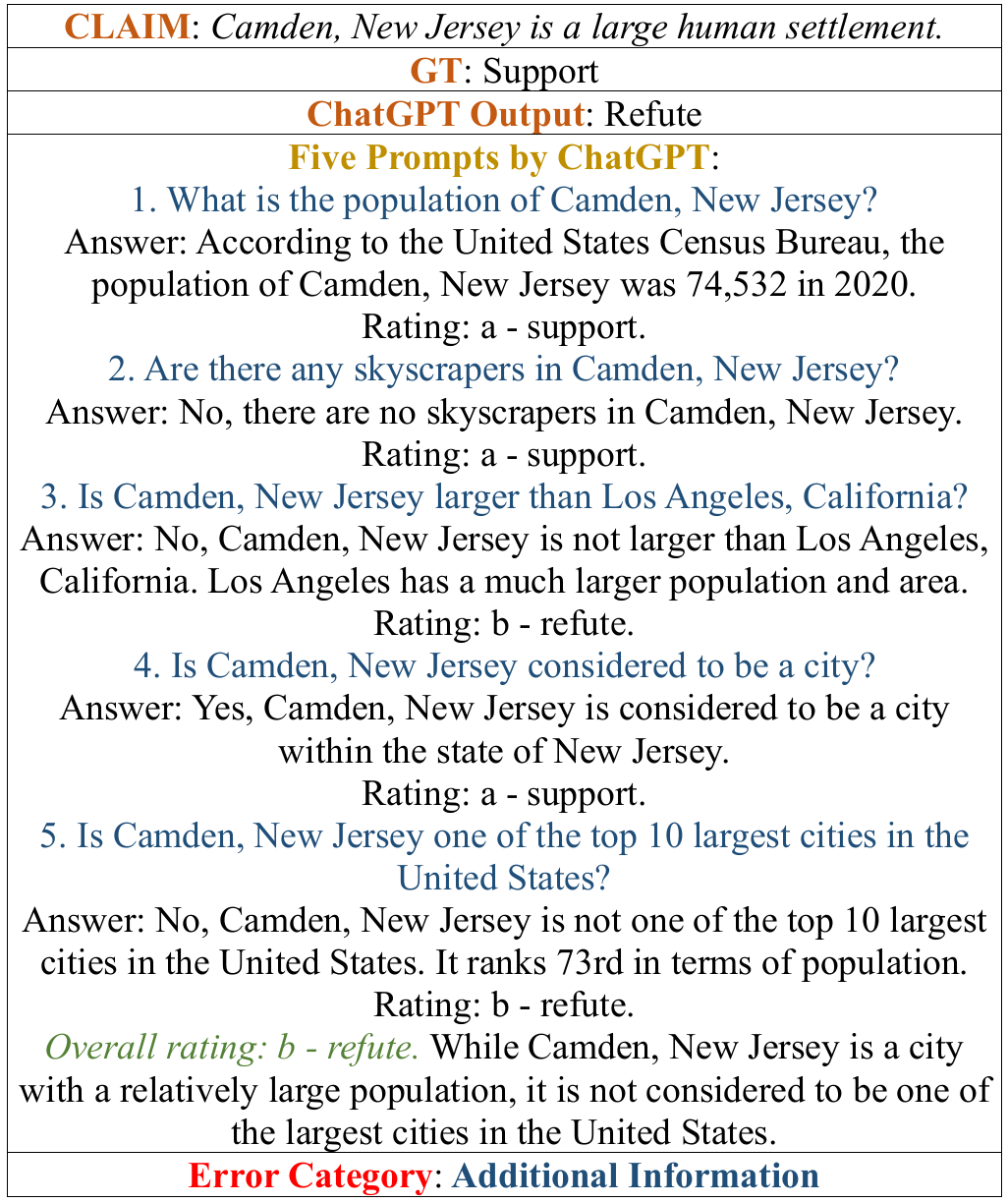}
    \caption{Additional Information}
    \label{fig:fig6}
\end{figure}

Figure \ref{fig:fig6} examines the error category \textit{Additional information} where ChatGPT made a mistake because of the availability of additional knowledge. Specifically for the claim "Camden, New Jersey is a large human settlement.", the response provided by ChatGPT compared the city with other cities in the United States, resulting in an erroneous response. The cause of the error is due to the availability of additional information about other cities in the United States to ChatGPT.

\begin{figure}
    \centering
    \includegraphics[width=0.49\textwidth]{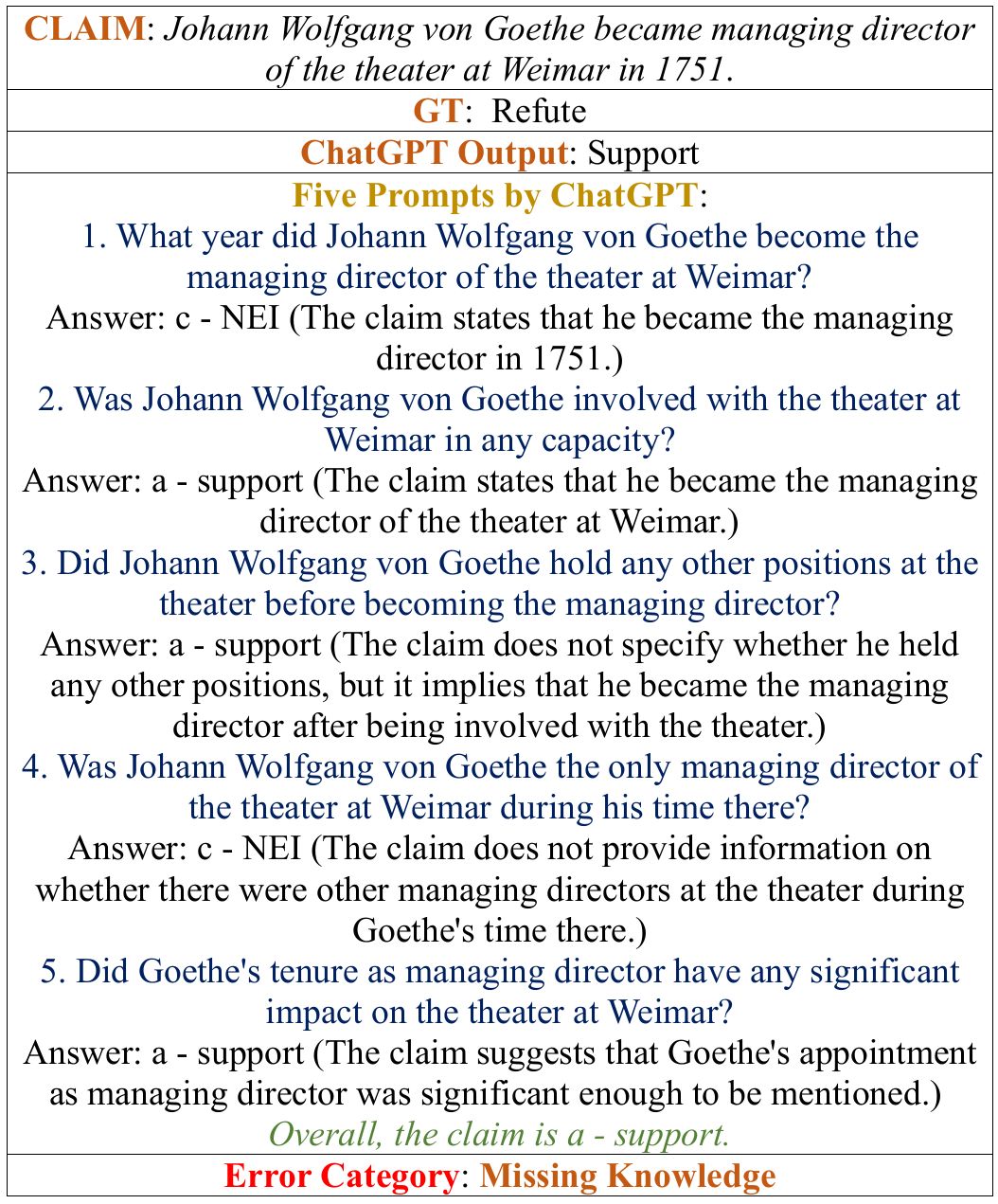}
    \caption{Missing Knowledge}
    \label{fig:fig7}
\end{figure}

Figure \ref{fig:fig7} examines the error category \textit{Missing knowledge}, which is caused due to missing knowledge related to the claim. The example shows that the answers of ChatGPT for the generated five prompts never mentioned the year when Johann Wolfgang von Goethe became the managing director of the theater at Weimar. Specifically, for the prompt, "What year did Johann Wolfgang von Goethe become the managing director of the theater at Weimar?", the response of ChatGPT is NEI implying missing knowledge.

\begin{figure}
    \centering
    \includegraphics[width=0.49\textwidth]{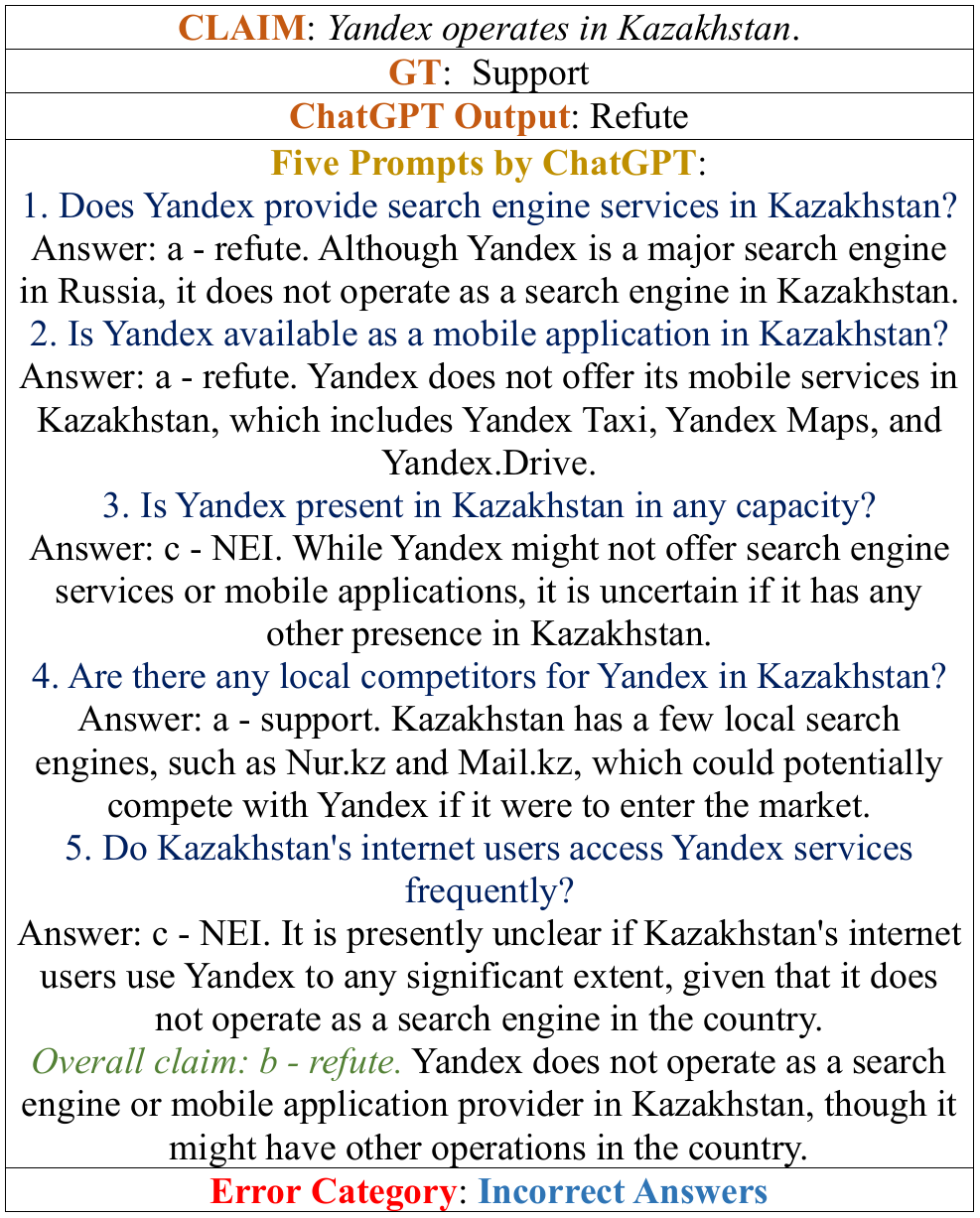}
    \caption{Incorrect Answers}
    \label{fig:fig8}
\end{figure}

Figure \ref{fig:fig8} examines the error category \textit{Incorrect answers}, where ChatGPT gave incorrect answers for non-factual prompts. From the example, we can observe that even though Yandex operates in Kazakhstan, all the responses by ChatGPT for the generated prompts suggest that Yandex does not operate in Kazakhstan, leading to an incorrect overall response.

\begin{figure}
    \centering
    \includegraphics[width=0.49\textwidth]{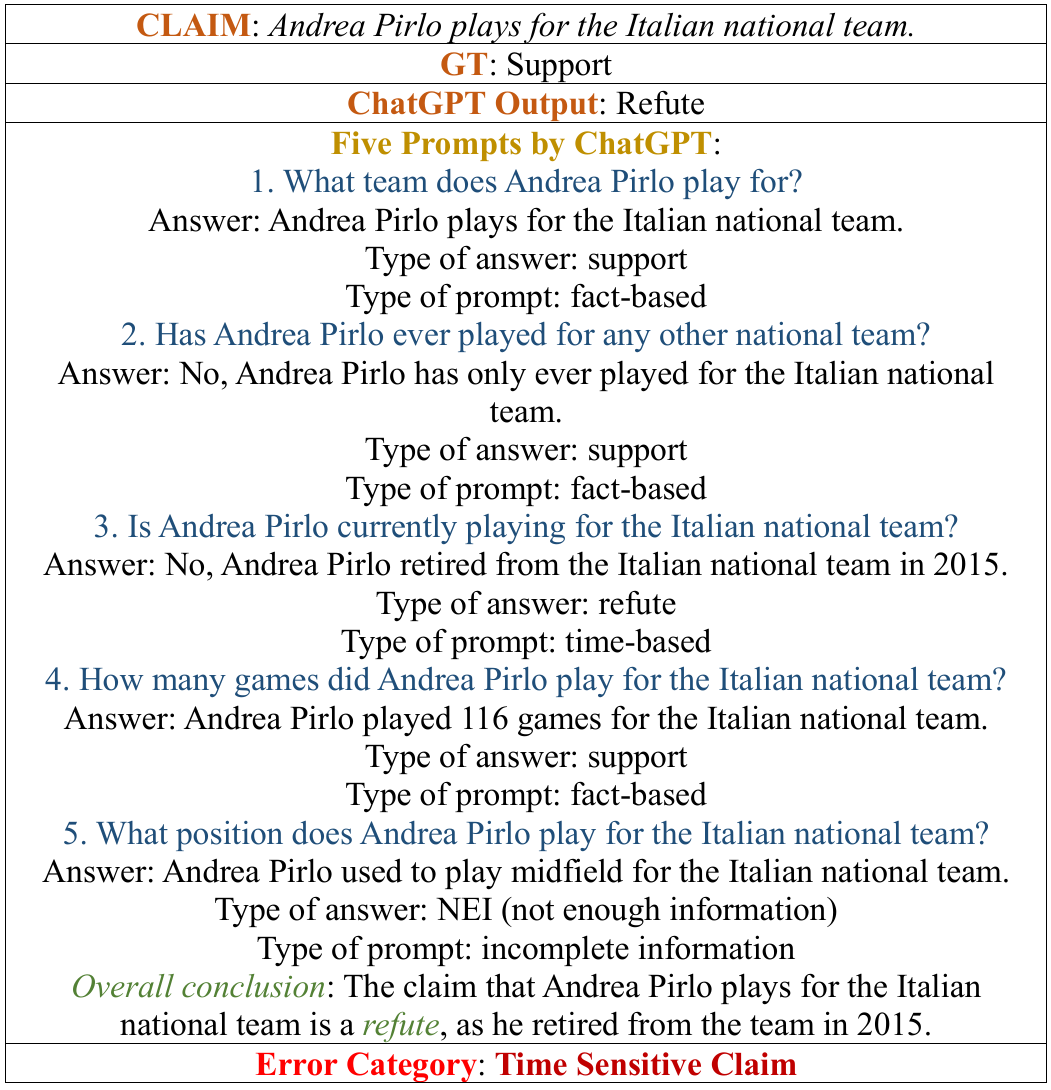}
    \caption{Time Sensitive Claim}
    \label{fig:fig9}
\end{figure}

Figure \ref{fig:fig9} discusses the error category  \textit{time sensitive claim} where the claim is time bounded. Andrea Pirlo played for the Italian National team and retired in 2015. ChatGPT correctly captured this time-sensitive information and generated the correct response. However, the response contradicts the ground truth of the static FEVER 1.0 shared task dataset leading to the error. However, the percentage of errors in this category is low, and addressing the evaluation to remove these false errors will not increase ChatGPT performance much.

\subsection{Case Study}
\label{sec:case_study}

We conduct a case study to show that the performance of ChatGPT is sensitive to different prompts. From the results in Table \ref{table: main_table}, we observe that \textit{Prompt-3} achieves significantly better performance compared to \textit{Prompt-1} and \textit{Prompt-2}. In this case study, we analyze a few examples where ChatGPT provided the correct response to the claim using \textit{Prompt-3} and provided an incorrect response when \textit{Prompt-1} and \textit{Prompt-2} are used. For the case study, we provide three examples of claims in Table \ref{tab:case_study} and \ref{tab:case_study1}. The examples cover three scenarios with ground truth labels as Refute, NEI, and Support. The case study validates our assertion that asking ChatGPT to generate multiple prompts and answer them helps ChatGPT make guided decisions since answering the prompts creates a flow similar to a conversation.

\begin{table*}[h]
    \centering
    \caption{Case study showing the performance improvement of ChatGPT on Prompt-3}
    \resizebox{2\columnwidth}{!}{%
    \begin{tabular}{p {5 cm}|p{2 cm} | p{2cm} | p{8cm} | p{8cm}}
    \hline
    \textbf{Claim} & \textbf{Ground Truth} & \textbf{Prompt-1} & \textbf{Prompt-2} & \textbf{Prompt-3}\\
    \hline
    Tenacious D started in 1997. & Refute & Support & Evidence: Tenacious D released their first album, self-titled "Tenacious D" in September 2001, but had been performing together since 1994 and recorded their first demo tape in 1997. Based on this evidence, the claim that "Tenacious D started in 1997" is a (a.) support - the evidence confirms that the band began playing together and recording music in 1997, even though their first album was not released until 2001. & 
    1. When did Tenacious D first perform live?
    
    Answer: Tenacious D performed live for the first time in 1994.
    
    Answer Type: Refute
    
    Claim Type: Refute

    2. What year was Tenacious D's debut album released? 
    
    Answer: Tenacious D's self-titled debut album was released in 2001.
    
    Answer Type: Refute
    
    Claim Type: Refute

    3. Did Tenacious D win a Grammy Award before 2000? 
    
    Answer: No, Tenacious D did not win a Grammy Award before 2000.
    
    Answer Type: Support
    
    Claim Type: Support

    4. When did Tenacious D release their first single? 
    
    Answer: Tenacious D released their first single "Tribute" in 2000.
    
    Answer Type: Refute
    
    Claim Type: Refute

    5. How old were Jack Black and Kyle Gass when they formed Tenacious D? 
    
    Answer: Jack Black and Kyle Gass were in their mid-20s when they formed Tenacious D in 1994.
    
    Answer Type: Support
    
    Claim Type: Support

    Overall claim: Refute\\ 
    \hline

    Tilda Swinton is a vegan. & NEI & Support & Evidence for the claim: In a 2019 interview with British Vogue, Tilda Swinton stated: "I'm a vegan, but I have to admit I'm not really that pure about it - it's a work in progress." Based on the evidence, the claim is (a.) - support. & 1. Has Tilda Swinton ever publicly stated that she follows a vegan diet?
Answer: a - Refute
Explanation: There is no evidence that Tilda Swinton has ever publicly confirmed that she follows a vegan diet.

2. Has Tilda Swinton ever been seen wearing or promoting animal products or clothing made of animal materials?
Answer: c - NEI
Explanation: There is no evidence to suggest that Tilda Swinton wears or promotes animal products or clothing made of animal materials, but this does not necessarily confirm or refute the claim that she is vegan.

3. Does Tilda Swinton actively support or donate to animal rights organizations?
Answer: a - Support
Explanation: Tilda Swinton has been known to actively support and participate in causes related to animal rights. In 2019, she attended a protest in London against the use of fur in the fashion industry.

4. Has Tilda Swinton ever mentioned the reasons behind her dietary choices in interviews or public appearances?
Answer: b - Refute 
Explanation: There is no evidence that Tilda Swinton has ever talked about her dietary choices in public, which makes it difficult to confirm or refute the claim that she is vegan.

5. Are there any documented instances where Tilda Swinton has knowingly consumed animal products?
Answer: b - Refute 
Explanation: There is no evidence that suggests that Tilda Swinton has knowingly consumed any animal products. In fact, she has been known to avoid meat and dairy products in her diet. 

Final verdict: Although there is no clear evidence to confirm that Tilda Swinton is vegan, she has actively supported animal rights organizations in the past and is known to exclude meat and dairy products from her diet. Based on the prompts and answers provided, the claim can be considered a - NEI.

Overall claim: NEI \\ \hline
    \end{tabular}
    }
    \label{tab:case_study}
\end{table*}

\begin{table*}[h]
    \centering
    \caption{Case study showing the performance improvement of ChatGPT on Prompt-3 (Contd.)}
    \resizebox{2\columnwidth}{!}{%
    \begin{tabular}{p {5 cm}|p{2 cm} | p{2cm} | p{8cm} | p{8cm}}
    \hline
    \textbf{Claim} & \textbf{Ground Truth} & \textbf{Prompt-1} & \textbf{Prompt-2} & \textbf{Prompt-3}\\
    \hline
    Rick Yune was on a series that cancelled on December 12. & Support & NEI & (b.) - Refute. 

As an AI language model, I do not have access to real-time information or current events, but upon researching output sources, I have found no evidence to support this claim. Therefore, I cannot confirm the claim but based on the lack of evidence found, it can be considered as refuted.
 & 1. Was Rick Yune a main cast member on a TV series that was cancelled on December 12?
Answer: a - refute; Rick Yune did not appear to be a main cast member on any TV series that was cancelled on December 12.

2. Did Rick Yune make a guest appearance on a TV series that was cancelled on December 12?
Answer: c - NEI; There is no clear evidence either confirming or refuting this claim.

3. Was Rick Yune involved in the production of a TV series that was cancelled on December 12?
Answer: a - support; Rick Yune was involved in the production of the TV series "Altered Carbon," which was cancelled on December 12.

4. Was Rick Yune's character written out of a TV series that was cancelled on December 12?
Answer: c - NEI; It is unclear whether or not Rick Yune's character was written out of any TV series that was cancelled on December 12.

5. Did Rick Yune express disappointment about the cancellation of a TV series on December 12?
Answer: b - refute; There is no evidence suggesting that Rick Yune expressed disappointment about the cancellation of any TV series on December 12 or any other date. 

Overall claim: Rick Yune was on a series that cancelled on December 12.
Answer: a - support; Rick Yune was involved in the production of the TV series "Altered Carbon," which was cancelled on December 12.
    \\ \hline
    \end{tabular}
    }
    \label{tab:case_study1}
\end{table*}

\end{document}